# Computer Interaction and the Benefits of Social Networking for People with Borderline Personality Disorder: Enlightening Mental Health Professionals

Alice Good, Arunasalam Sambhanthan, Vahid Panjganj, Samuel Spettigue

School of Computing
University of Portsmouth
PO1 3AE
UK

alice.good@port.ac.uk

**Abstract.** This paper seeks to present the findings of a focus group and questionnaire in assessing how aware mental health professionals, who have experience with people with Borderline Personality Disorder (BPD), are in the extent of ICT based support for people with BPD. The methods used were both qualitative and quantitative and used descriptive data. Content analysis was used to explore specific themes and results were cross-examined between the two methods. The work should be viewed as an exploratory study into the viability and likely acceptance of a virtual support community specifically designed for people with BPD. The long term aim is to provide additional support for people with BPD, especially when they are in crisis and might be at a higher risk of harm.

## 1 Introduction

The purpose of this research is to examine the level of awareness that mental health professionals have of ICT based support for people with BPD. The research presented here questions professionals on their awareness of Facebook support groups and other Web based resources. It also looks at how likely these professionals would accept a novel proposal for a virtual therapeutic type support environment, specifically for people with BPD.

Treatment for severe mental illness does not often consider the potential that social networking can have in both reducing the impact of loneliness (Perese & Wolf, 2005), and the sense of 'feeling alone' (Neal & McKenzie, 2010), and so the use of ICTs in providing support for people with BPD could certainly be beneficial. People with Borderline Personality Disorder (BPD) represent 0.7% of the UK population (NICE, 2009) and are reported as more likely to seek psychiatric intervention than those with other psychiatric disorders (Rendu et al., 2002, cited in NICE, 2009). Whilst there is ICT based support specifically for people with BPD, this is limited predominantly to discussion forums, emails and social networking sites facilitated by Face Book.

Second Life virtual world communities could be beneficial in providing support for people with BPD who might be experiencing a sense of 'feeling alone', and who wish to seek support and useful information on resources available. Second Life, with over 15 million users (K Zero Universe, 2009) is growing as a medium for social interaction. A recent survey on healthcare related activities using Second Life shows that patient education and awareness building as the major health related activity undertaken. Mental health groups in Second Life featured the largest number of members at 32% of the total users (Norris, 2009). In terms of categories of groups, 15% of the health support groups in Second Life were dedicated to mental health. Second Life could then potentially offer the means to provide a virtual support system for people with BPD.

## 2 Method

The methodology applied to this research comprises of two methods; a focus group and a follow up questionnaire to validate and supplement the data. Both qualitative and quantitative data were sought from the study. Participants included nineteen psychiatric staff, some who work as part of an outreach team and others as registered mental nurses in hospitals. All were professionals who had experience of working with people with BPD, most of whom were currently supporting and/or treating current patients with a BPD diagnosis. Both methods included the following themes: types of treatment and support provided; the need for additional support; the extent that staffs were aware of ICT support for people with BPD and how likely they were to accept and promote a virtual support environment. The data collected from the focus group was subjected to content analysis, which was then cross referenced against related questions within the questionnaire.

## 3 Results

Participants comprising of mental health professionals from the focus group were questioned on the type and extent of support they provided for people with BPD. Responses ranged from pharmaceutical treatment, referrals to Dialectical behavior therapy (DBT) and simply 'being there' to listen and support.. All participants reported that the degree of time spent supporting people with BPD was significantly higher than their other clients. This was further elaborated on with one member of the group stating :

*"I think people (professionals) tend to shy away from it. I don't think they know what to do with people with BPD."*

Other members confirmed they agreed. The questionnaires results provided further information, with all participants stating that in their experience, people with BPD require high levels of support. Furthermore, there were opinions expressed both in the focus group and the questionnaire that people with BPD were 'untreatable'.

In spite of this evident need for high levels of support, professionals did not actively promote any types of self help forms of methods via the Internet. See figure 1. The results elicited from the questionnaire further support this.

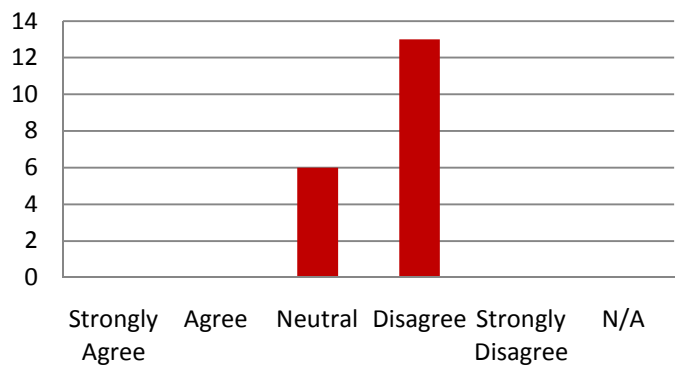

**Figure 1: As a professional working with people with BPD, I encourage the use of specific self help methods via the Internet.**

The results from both methods showed that professionals felt that people with BPD were keen to meet others who shared the same diagnosis to meet and support one another. Results from this study have already indicated that the mental health professionals questioned did not promote self help methods via the Internet. When questioned on how aware they were on the availability of online support and resources specifically for people with BPD, including Facebook support groups and information sites where specific examples were given, none of the focus group participants conveyed any awareness of their existence. These results were further validated via a follow-up questionnaire. See figure 2.

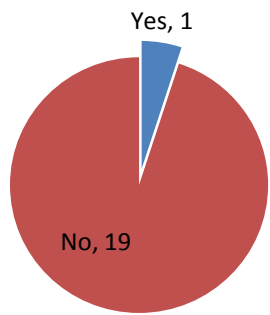

**Figure 2: Awareness of mental health professionals of Web based support groups and resources for people with BPD**

Whilst these professionals are evidently less aware of ICT based support, they were able to give example of offline resources such as the Steps Programme (A.A.), which is used to provide support to people with addiction and other behavioral problems.

In continuing with theme of the need for additional support for people with BPD, one participant, representing outreach support workers stated:

*"If they have the ongoing support, over the computer 24/7, it could be useful. We only work 9 - 4.30 and no out of hours service. So we point people to Samaritans of Mental Health Line. In terms of an out of hours crisis, we can't provide any support. Because that's the thing isn't it, they need support most of the time."*

These results indicate that there is a clear need for additional support for people with BPD. While there are support groups available online. including those on Facebook, which could offer a degree of containment and/or support, there was no awareness of their existence. In spite of this lack of awareness, there is a keenness to promote any additional support available that might assist people with BPD. Moreover, when the idea was put forward of a virtual support environment, specifically catering for people with BPD, participants responded favourably.

*"Anything that would offer extra support for people with BPD, we would embrace."*

Furthermore, all participants expressed an interest in being involved in a pilot study with the aim to research how a virtual support community system could be designed for people with BPD.

## 4. Discussion and Conclusions

This research is limited in that only a relatively small sample of mental health professionals were involved. However, the results do suggest that these professionals feel that people with BPD have a higher need for support than their other clients. Whilst there is great potential for the use of social networking support groups for people with mental health problems, including Facebook and Second life, these professionals were not aware of their existence. In fact, this paper has highlighted a distinct lack of awareness from these professionals, of any online support specifically related to people with BPD. In spite of this lack of awareness, they were keen to promote any additional ICT related support that might benefit people with BPD and demonstrated a positive response to the proposal of a potential virtual therapeutic support community, specifically designed for people with BPD.

Reviewed literature reports that social networking and second life are popular methods for the provision of support for people with general mental health problems

(Perese & Wolf, 2005, Norris, 2009). With the economic impact that people with BPD place upon health resources, particularly in comparison to other service users (Rendu et al., 2002, cited in NICE, 2009), more awareness of ICT support would be useful to professionals for them to be able to advise their clients to utilise them.

Whilst the advent of social networking has given rise to online support groups for people with BPD, there is also potential scope to utilise resources such as second life in providing further support. Furthermore, there is some potential that the provision of additional support could have some impact upon reducing emergency hospital admissions for people with BPD, as well as a possibility for the subsequent decrease in the economic impact to the health services.

This work is a continuation of the research carried out in exploring the potential of virtual therapeutic communities based on existing models of therapeutic hospitals and communities, as well as virtual treatments and support in treating people with BPD. An interdisciplinary approach to this research features collaboration from experts in HCI, forensic psychology and psychotherapy. Further studies will focus on developing a framework for the design of a virtual TC environment, which will be very much a user driven study, featuring mental health professionals and people with a diagnosis of BPD. Furthermore, research into exploring how effective avatars can be in enabling a sense of distancing when 'off loading' emotions and thoughts will be carried out. The proposed environment could potentially be used as a framework to support other mental health disorders. It should be emphasised that this proposal is intended to provide additional support and is in no way proposed as a replacement to existing methods of support and treatment.

Following the research indicated here, there is a need for improved awareness of the availability of current online resources and support. A user-centred approach to the design of a virtual support environment specifically for people with BPD, will certainly go some way in disseminating the idea. However to ensure end user acceptance, awareness and promotion by key professionals will be crucial.